\documentclass[12pt,a4paper]{article}

\usepackage[latin2]{inputenc}
\usepackage[T1]{fontenc}

\usepackage{pslatex}
\usepackage{ae,aecompl}
\usepackage{epsfig}
\usepackage{graphics}
\usepackage{graphicx}
\usepackage{amsmath}
\usepackage{amssymb}
\usepackage{pifont}
\usepackage{psfrag}

\usepackage{graphicx}
\usepackage{dcolumn}
\usepackage{bm}

 \def\ccc#1;#2{\left\langle #1
\left\vert #2 \right.\right\rangle} 

\begin{document}

\title{On the origin of the Epps effect}

\author{Bence T\'oth\footnote{E-mail: bence@maxwell.phy.bme.hu} $^{1,2}$ \and  J\'anos Kert\'esz$^{2,3}$}

\maketitle

$^1$ ISI Foundation - Viale S. Severo, 65 - I-10133 Torino, Italy 

$^2$ Department of Theoretical Physics, Budapest University of Technology and 
Economics - Budafoki \'ut. 8. H-1111 Budapest, Hungary

$^3$ Laboratory of Computational Engineering, Helsinki University of Technology 
-  P.O.Box 9203, FI-02015, Finland

\begin{abstract}
The Epps effect, the decrease of correlations between stock returns 
for short time windows, was traced back to the trading asynchronicity 
and to the occasional lead-lag relation between the prices. We study pairs 
of stocks where the latter is negligible and confirm the importance of 
asynchronicity but point out that alone these aspects are insufficient to give 
account for the whole effect.
\end{abstract}

\section{Introduction}
1979 T. W. Epps reported empirical results showing that stock return 
correlations decrease as the sampling frequency increases \cite{epps}. Later 
this phenomenon has been found in several studies of different stock markets 
\cite{bonanno2001,zebedee2001} and of foreign exchange markets 
\cite{lundin1999,muthuswamy2001}. Since the most important factors in classical 
portfolio management are the correlations between the individual assets, the 
understanding and the accurate description of these correlations on all time 
scales are of major importance.

Considerable effort has been devoted to uncover the relationship of the 
correlation coefficient and the sampling frequency 
\cite{reno,iori2004,iori2006,takayasu2003,kwapien2004,zhang2006}. So far the 
studies have revealed two factors causing the Epps effect. The first one is the 
lead-lag effect \cite{lo,kullmann,toth2006} between the stock returns, which can 
occur for pairs of stocks of very different capitalization and/or for some 
functional dependencies between them. In this case the maximum of the 
time-dependent correlation function is at non zero time lag  resulting in 
increasing correlations as the sampling time scale gets in the same order of 
magnitude as the characteristic lag (typically of the order of minutes) 
\cite{zebedee2001,reno}. However, in a recent study \cite{toth2006} we showed 
that through the years this effect becomes less important since the 
characteristic lag shrinks, signalizing an increasing market efficiency. 

The second, more important factor is the asynchronicity of 
trading \cite{reno,iori2004,zhang2006,lo2}. Empirical results showed that  
taking into account only synchronous price ticks the Epps effect is reduced, 
i.e., the measured correlations increase. It is natural to assume that, for a 
given sampling frequency, increasing trading activity should enhance 
synchronicity, leading to a weaker Epps effect. Indeed, using Monte Carlo 
experiments, an inverse relation was found between trading activity and the 
correlation drop \cite{reno}.

The aim of the present communication is to investigate empirically the role of 
the trading activity on the Epps effect. We find that in contrast to the 
simulations, the empirical data do not scale in the expected way indicating that 
a further, possibly human factor is also at play.
\section{Methodology}
\label{meth}

In our analysis we used the Trade and Quote (TAQ) Database of the New York Stock 
Exchange (NYSE) for the period of 4.1.1993 to 31.12.2003, containing tick-by-tick 
data. To avoid the problems occurring from splits in the prices of stocks, which
cause large logarithmic return values in the time series, we applied a
filtering procedure. In high-frequency data, we omitted returns larger
than 5\% of the current price of the stock. This retains all
logarithmic returns caused by simple changes in prices but excludes
splits which are usually half or one third of the price. We computed correlations 
for each day separately and averaged over the set of days, this way avoiding large 
overnight returns and trades out of the market opening hours.

We computed 
the logarithmic returns of stock prices:
\begin{eqnarray}
\label{ret}
r_{\Delta t}^{A}(t)=\ln \frac{p^{A}(t)}{p^{A}(t-\Delta t)},
\end{eqnarray}

where \(p^{A}(t)\) stands for the price of stock \textit{A} at time \(t\). The 
time dependent correlation function \(C_{\Delta t}^{A,B}(\tau)\) of stocks 
\textit{A} and \textit{B} is defined by

\begin{eqnarray}
\label{C}
C_{\Delta t}^{A,B}(\tau)=\frac{\left\langle r_{\Delta t}^{A}(t)r_{\Delta 
t}^{B}(t+\tau)\right\rangle - \left\langle r_{\Delta t}^{A}(t)\right\rangle 
\left\langle r_{\Delta 
t}^{B}(t+\tau)\right\rangle}{\sigma^{A}\sigma^{B}}.\end{eqnarray}

The notion 
\(\left\langle \cdots\right\rangle\) stands for the moving time average over the 
considered period:
\begin{eqnarray}\label{time_ave}
\left\langle r_{\Delta t}(t)\right\rangle =\frac{1}{T-\Delta t}\sum_{i=\Delta t}^{T} r_{\Delta t}(i),
\end{eqnarray}
where time is measured in seconds and T is the time span of the data.

\(\sigma\) is the standard deviation of the return 
series:
\begin{eqnarray}\label{sigma}\sigma=\sqrt{\left\langle r_{\Delta 
t}(t)^{2}\right\rangle - \left\langle r_{\Delta 
t}(t)\right\rangle^2}.
\end{eqnarray}

For pairs of stocks with a lead--lag effect the function $C_{\Delta 
t}^{A,B}$ has a peak at non-zero $\tau$. 

The equal-time correlation coefficient is: \(\rho_{\Delta t}^{A,B}\equiv
C_{\Delta t}^{A,B}(\tau=0)\).In our notations the Epps effect means that
\(\rho_{\Delta t}\) decreases as \(\Delta t\) decreases (see Figure
\ref{fig1}).The prices are defined as being constant between two consecutive
trades (previous tick estimator), thus the \(\Delta t\) time scale of the
sampling can be chosen arbitrarily.

\section{Results}\label{results}

In order to separate the different origins of the Epps effect we consider only 
pairs of stocks where the lead--lag effect is neglible, i.e., for which the 
price changes are highly correlated with the peak position of $C_{\Delta 
t}^{A,B}$ of Eq. \ref{C} being at $\tau \approx 0$. 
Good candidates are the pairs Coca-Cola Company (KO) / PepsiCo, Inc. (PEP), Caterpillar Inc. (CAT) / Deere \& Company (DE), Wal-Mart Stores, Inc. (WMT) / Sprint Nextel Corporation (S), etc.
We shall illustrate our findings on the example of KO/PEP but the other 
pairs show similar behaviour.

\begin{figure}  \centering 
\includegraphics[width=0.50\textwidth]{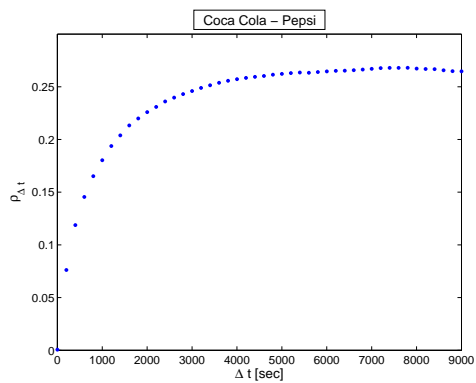}   \caption{The 
correlation coefficient as a function of sampling time scale for the period 
1993--2003 for the Coca-Cola Pepsi pair. Several hours are needed for the 
correlation to reach it's asymptotic value.}  \label{fig1}
\end{figure}

In Figure \ref{fig1} we show the correlation coefficient as a function of 
the sampling time scale for the whole period, 1993--2003.One can see that the 
correlations increase as the sampling frequency decreases. The growth is very 
fast in the beginning but several hours are needed for the correlation to reach 
its asymptotic value.

\begin{figure}  \centering              
\includegraphics[width=0.50\textwidth]{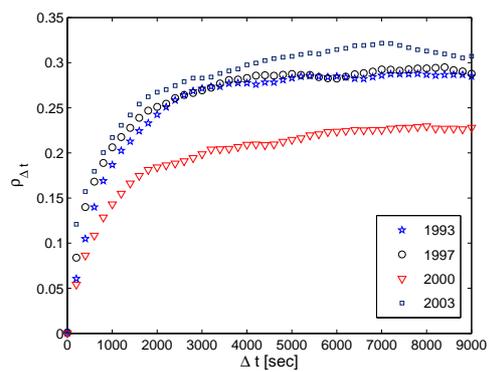}        \caption{The 
correlation coefficient as a function of sampling time scale for the years 1993, 
1997, 2000 and 2003 for the Coca-Cola Pepsi pair.}      
\label{fig2}\end{figure}

To study 
the trading frequency dependence of the correlation drop we computed the 
Epps-curve separately for different years. Figure \ref{fig2} shows the 
correlation coefficient as a function of the sampling time scale for the years 
1993, 1997, 2000 and 2003.

\begin{figure}  
\centering              
\includegraphics[width=0.50\textwidth]{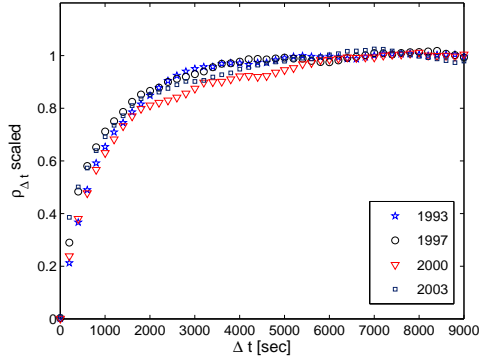} 
\caption{The correlation coefficient scaled with the asymptotic value as a 
function of sampling time scale for the years 1993, 1997, 2000 and 2003 for the 
Coca-Cola Pepsi pair. The scaled curves give a reasonable data collapse in 
spite of the considerably changing trading frequency (see Fig \ref{fig4}).}     
\label{fig3}\end{figure}

As it 
is known, correlations are not constant through the years. Apart from a growing 
trend, being a sign of growing efficiency \cite{toth2006}, there are 
fluctuations which depend on the whole market. For example in crash periods the 
complete market is moving together. We have to take this into account and try to 
extract the effect of changing asymptotic correlations from the phenomenon 
studied. In order to do this, we scaled the curves with their asymptotic value: 
The latter was defined as the mean of the correlation coefficients for the 
sampling time scales \(\Delta t=6000\) seconds through \(\Delta t=9000\) 
seconds, and the correlations were divided by this value. Figure \ref{fig3} 
shows the scaled curves for the years 1993, 1997, 2000 and 2003. 

\begin{figure}  \centering              
\includegraphics[width=0.50\textwidth]{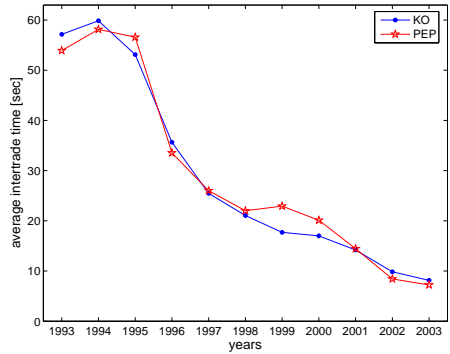}  \caption{The 
average inter-trade time for the years 1993 to 2003 for the stocks of The 
Coca-Cola Company (KO) and PepsiCo, Inc. (PEP). The activity was growing almost 
monotonically.}  \label{fig4}\end{figure}

Knowing that the trading activity almost monotonically 
grew in the period studied (as it can be seen in Figure \ref{fig4}), one would 
expect the diminution of the Epps effect, and thus much weaker decrease of the 
correlations as sampling frequency is increased. However, after 
scaling with the asymptotic correlation value, the curves give a reasonable 
data collapse and no systematic trend can be seen. Surprisingly, increasing the 
trading frequency by a factor of $\sim 5$ does not lead to a measurable 
reduction of the characteristic time of the Epps effect.

These observations show that the effect of 
asynchronicity in trading is more complex than it is usually understood and can 
not be solely discussed through the trading activity. The characteristic time of 
the Epps effect seems to be independent of the trading frequency, indicating 
another, possibly human time scale being also at play. This assumption, together 
with an analytic treatment of the Epps effect and with Monte Carlo experiments 
will be investigated in a forthcoming paper \cite{toth2007}, which is based on a 
relation between equal time and time lagged correlations. A similar approach was 
used in \cite{bouchaud2005}.

\section*{Acknowledgments}
Support by OTKA T049238 is acknowledged.


\begin{thebibliography}{99}\bibitem{epps}T.W. Epps, Journal of the 
American Statistical Association \textbf{74}, 291-298 
(1979)\bibitem{bonanno2001}G. Bonanno, F. Lillo, R.N. Mantegna, Quantitative 
Finance \textbf{1}, 1-9 (2001)\bibitem{zebedee2001}A. Zebedee, A closer look at 
co-movements among stock returns, San Diego State University, \textit{working 
paper} (2001)\bibitem{lundin1999}M. Lundin, M. Dacorogna, U. A. M\"uller, 
Correlation of high-frequency financial time series. In P. Lequeux 
(Ed.),\textit{Financial Markets Tick by Tick}. Wiley \& Sons. 
\bibitem{muthuswamy2001}J. Muthuswamy, S. Sarkar, A. Low, E. Terry, Journal of 
Futures Markets \textbf{21}(2), 127-144 (2001)\bibitem{reno}R. Ren\`o, 
International Journal of Theoretical and Applied Finance \textbf{6}(1), 87-102 
(2003)\bibitem{iori2004}O. V. Precup, G. Iori, Physica A \textbf{344}, 252-256 
(2004)\bibitem{iori2006}O. V. Precup, G. Iori, European Journal of Finance 
(2006).\bibitem{takayasu2003}T. Mizuno, S. Kurihara, M. Takayasu, H. Takayasu, 
cond-mat/0303306 (March 2003)\bibitem{kwapien2004}J. Kwapie\'n, S. Dro\.zd\.z, 
J. Speth, Physica A \textbf{337}, 231-242 (2004)
\bibitem{zhang2006}
L. Zhang, Estimating Covariation: Epps Effect, Microstructure Noise 
\textit{working paper} (2006)  
\bibitem{lo}
A. Lo, A. C. MacKinlay, Rev. Finance Stud \textbf{3}, 175-205 (1990)

\bibitem{kullmann}
L. Kullmann, J. Kert\'esz, K. Kaski, Phys. Rev. E \textbf{66}, 026125 (2002)

\bibitem{toth2006}
B. T\'oth, J. Kert\'esz, Physica A \textbf{360} 505-515 (2006)

\bibitem{lo2}
A. Lo, A. C. MacKinlay, Journal of Econometrics \textbf{45}, 181-211 (1990)

\bibitem{toth2007}
B. T\'oth, J. Kert\'esz, in preparation 

\bibitem{bouchaud2005}
M. Potters, J.-P. Bouchaud, L. Laloux, Acta Physica Polonica B, \textbf{36}, 9, (2005)

\end{thebibliography}
\end{document}